\documentclass[prl,twocolumn]{revtex4}

\usepackage{graphicx}
\begin{document}

\title{Enhanced low-energy spin dynamics with diffusive character in the iron-based superconductor (La$_{0.87}$Ca$_{0.13}$)FePO: Analogy with high $T_c$ cuprates}

\author{Marc-Henri \textsc{Julien}$^{1}$ \footnote{E-mail: marc-henri.julien@ujf-grenoble.fr}}

\affiliation{$^{1}$Laboratoire de Spectrom\'etrie Physique, UMR5588 CNRS and Universit\'e J. Fourier - Grenoble, 38402 Saint Martin
d'H\`{e}res, France}

%\kword{NMR, superconductivity, LaCaFePO, iron-based oxypnictide}

\maketitle

The recent discovery of superconductivity above 20~K in iron-based layered materials \cite{Kamihara} has given rise to an avalanche of
experimental studies, reminiscent of the early days of cuprate superconductors. In a recent paper~\cite{Nakai}, Nakai and coworkers
reported on the NMR investigation of a related superconductor (La$_{0.87}$Ca$_{0.13}$)FePO. This phosphorus-containing material has a
$T_c$ of 6~K, that is much lower than its arsenide-containing isostructural counterpart LaFeAsO. At odd with studies in LaFeAsO, Nakai
{\it et al.} do not observed any signature of superconductivity in the NMR measurements. In particular, the spin-lattice relaxation rate
of $^{31}$P nuclei divided by temperature, $(T_1T)^{-1}$, shows a $\omega_{\rm NMR}$-dependent increase on cooling below $T_c$
($\omega_{\rm NMR}$ is the NMR frequency). The authors indicate that to the best of their knowledge such increased low-energy spin
dynamics have never been reported in the superconducting state and they suggest that these novel fluctuations could originate from a
spin-triplet symmetry of the superconducting state.

In this short note, we show that the data of Nakai {\it et al.} actually bears strong similarity with two previous observations in cuprate
superconductors: the divergence of the spin-spin correlation function at low frequency in Tl$_2$Ba$_2$CuO$_y$, and the absence of a
superconducting-like response in the NMR measurements of La$_{2-x}$Sr$_x$CuO$_4$ with $0.06\leq x \leq 0.12$.

\begin{figure}[t!]
%\vspace{-1cm}
\centerline{\includegraphics[width=9cm]{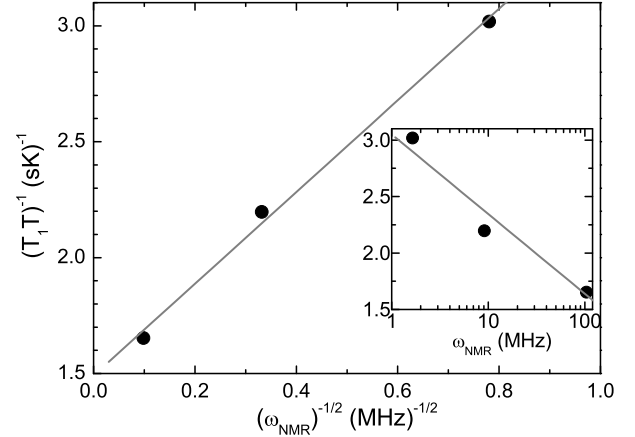}} %%%%%%%%%%%%%%%
%%%%%%%%%%%%%%%%%%%%%%%%%%%%%%%%
\vspace{-0.2cm}
%%%%%%%%%%%%%%%%%%%%%%%%%%%%%%%%%%%%%%%
\caption{\label{Fig1} Values of $(T_1T)^{-1}$ taken at $T=1.5$~K, as a function of $\omega_{\rm NMR}$ (Main panel) and a function of
$\log(\omega_{\rm NMR})$ (Inset). Data are taken from Nakai {\it et al.}~\cite{Nakai} with $\omega_{\rm NMR}=(1+K)\gamma H_0$, $K=0.3$\%,
$\gamma=17.24$~MHz~T$^{-1}$ for $^{31}$P and $H_0$=0.95, 5.25 and 60 kOe. Lines are linear fits to the data points.} \vspace*{-0.4cm}
\end{figure}

In \mbox{Fig. \ref{Fig1}}, we have plotted the value of $(T_1T)^{-1}$ measured at $T= 1.5$~K, as a function of $\omega_{\rm NMR}$, as
extracted from the Fig.~5 of Nakai {\it et al.}~\cite{Nakai}. $1/T_1$ clearly appears to be proportional to $\omega_{\rm NMR}^{-1/2}$, a
typical dependence for electronic spin diffusion in 1D systems, although the two-dimensional form $(T_1T)^{-1}\propto \ln(\omega^{-1})$
cannot be fully excluded at this stage (see the inset to Fig.~1). Spin diffusion corresponds to a situation where the spin-spin
correlation function has an anomalous tail at long times ({\it i.e.} a divergence at low $\omega$)~\cite{Boucher}. The $\omega_{\rm
NMR}^{-1/2}$ divergence of low-energy spin fluctuations has been observed in one-dimensional spin chains (see references contained in
recent works \cite{diffusion1D}).

%%%%%%%%%%%%%%%%%%%%%%%%%%%%%%%%%%%%%%%
The results of Nakai {\it et al.} are also reminiscent of a NMR study of Tl$_2$Ba$_2$CuO$_y$ by Kambe and coworkers~\cite{Kambe}. In this
high $T_c$ cuprate, the occurrence of 2D spin diffusion was deduced from the experimental observation $(T_1T)^{-1}\propto
\ln(\omega^{-1})$, for both $^{205}$Tl and $^{63}$Cu nuclei, {\it even above $T_c$}. We note that, to the best of our knowledge, the exact
nature of this diffusive process in a cuprate superconductor has not been elucidated yet.

Why 1D, rather than 2D, spin diffusion would be observed in the layered (La$_{0.87}$Ca$_{0.13}$)FePO is unclear, although this results
immediately raises questions on the fascinating possibility of a (yet-undetected) symmetry breaking in the electronic ground state. It
would be interesting to perform similar frequency dependent $T_1$ measurements in FeAs superconductors. The striped character of their
(undoped) spin density wave state may evolve into quasi one-dimensional electronic correlations upon doping with charge carriers.

It is also worth considering another analogous situation: $^{139}$La $T_1$ measurements in La$_{2-x}$Sr$_x$ CuO$_4$, with $0.06\leq
x\leq0.12$ also show an enhancement on cooling throughout the superconducting state. This increase is due to the freezing of magnetic
fluctuations on approaching the cluster spin glass transition~\cite{Julien}. In this case also, the occurrence of superconductivity is not
detected in the NMR data. In La$_{2-x}$Sr$_x$CuO$_4$, the onset for the enhancement of $1/T_1$ does not usually coincide with the
superconducting $T_c$, but both occur in a similar temperature range. It is thus also possible that (La$_{0.87}$Ca$_{0.13}$)FePO is close
to a magnetic instability and that the enhancement of $1/T_1$ occurs close to the zero-field $T_c$ by chance. If magnetism is nucleated
inside the vortex cores~\cite{Lake}, the enhancement of $T_1$ could even start at $T_c$.

In conclusion, (La$_{0.87}$Ca$_{0.13}$)FePO displays an anomalous enhancement of spin excitations, with possible diffusive character, in
its superconducting state. This findings points to an interesting similarity with some of the cuprate superconductors. The work of Nakai
and coworkers should thus stimulate theoretical attention as well as more detailed NMR measurements at low $T$ as a function of the
frequency in these promising Fe-based superconductors.

\section*{Acknowledgment}
I wish to thank K. Ishida, C. Berthier, P. Carretta and H. Mayaffre for useful exchanges.

\end{document}